\documentstyle[12pt]{article}

\newcommand{\F}{\noindent}

\newcommand{\SP}{\smallskip}
\newcommand{\MP}{\medskip}
\newcommand{\BP}{\bigskip}
\newcommand{\HH}{{\cal H}}

\begin{document}
\rightline{KIMS-1997-08-22}
\rightline{gr-qc/9708055}
\BP

\vskip2pt

%namelist environment (Nelson H. F. Beebe)
%form: \begin{namelist}{width}
\newcommand{\namelistlabel}[1]{\mbox{#1}\hfil}
\newenvironment{namelist}[1]{%
\begin{list}{}
{\let\makelabel\namelistlabel
\settowidth{\labelwidth}{#1}
\setlength{\leftmargin}{1.1\labelwidth}}
}{%
\end{list}}

\vskip6pt

\begin{center}

\Large

{\bf Comments on the Problem of Time}
\vskip6pt

\normalsize

\BP

\vskip10pt

Hitoshi Kitada{\footnote[2]{On leave from University of Tokyo}}

\SP

Department of Mathematics

University of Virginia

Charlottesville, VA 22903, USA

e-mail: kitada@ms.u-tokyo.ac.jp

home page: http://kims.ms.u-tokyo.ac.jp/

and

Lancelot R. Fletcher

\SP

The Free Lance Academy

30-3406 Newport Parkway

Jersey City, New Jersey 07310, USA

e-mail: lance.fletcher@freelance.com

home page: http://www.freelance.com/

\end{center}

\vskip6pt

\MP

\leftskip=24pt
\rightskip=24pt

\small

\F
{\bf Abstract.} 
The problem of time, considered as a problem
in the usual physical context, is reflected
in relation with the paper by Kauffman and
Smolin (\cite{KS}). It is shown that
the problem is a misposed problem in the
sense that it was raised with a lack of
the recognition of mathematically known facts.

\MP

\leftskip=0pt
\rightskip=0pt

\normalsize

\vskip 8pt

\F
We found an idea on time similar to ours
 in the paper ``A possible solution to
 the problem of time in quantum cosmology" (\cite{KS})
 by Kauffman and Smolin.

Their argument concerning the nature of time seems to converge
 with our understanding of time as local notion
 (\cite{K1}, \cite{K2}, \cite{K3}, \cite{Ki-Fl}),
 even though we come at this issue from
 completely different angles.

That their idea might lead to our notion of local time can be seen
especially from the summary passage in \cite{KS}, near the end
of the section entitled ``Can we do physics without a constructible
state space?":

They offer a way of interpreting the evolution of quantum states in
terms of a set of discrete, finite spin networks, each member of
which has a successor network. They argue that by focusing attention
on finite successor states it is possible to construct the relevant
(local) probability amplitudes without constructing the total Hilbert
space. They write:

\begin{quote}
        The theory never has to ask about the whole space of states, it
        only explores a finite set of successor states at each step.
	...
\end{quote}
\begin{quote}
        The role of the space of all states is replaced by the notion of
        the successor states of a given network .... They are finite in
        number and constructible. They replace the idealization of all
        possible states that is used in ordinary quantum mechanics.
	...
\end{quote}
\begin{quote}
        In such a formulation there is no need to construct the state
        space a priori, or equip it with a structure such as an inner
        product. One has simply a set of rules by which a set of
        possible configurations and histories of the universe is
        constructed by a finite procedure, given any initial state. In a
        sense it may be said that the system is constructing the space
        of its possible states and histories as it evolves.
        
        Of course, were we to do this for all initial states, we would
        have constructed the entire state space of the theory. But there
        are an infinite number of possible initial states and, as we
        have been arguing, they may not be classifiable. In this case it
        is the evolution itself that constructs the subspace of the
        space of states that is needed to describe the possible futures
        of any given state. And by doing so the construction gives us an
        intrinsic notion of time.
\end{quote}
In these passages, Kauffman and Smolin draw two particular
conclusions about the nature of time, both of which correspond to
the main points of our theory of local times:
(a) that time is essentially local, and (b) that time is to be
identified with the evolution of the system itself (which we call
the ``local system" in the following to indicate that it is a
subsystem of the total system, and we call the time of the local
system a local time of the local system.)
\BP

We now wish to make some comments on the proposal that
 Kauffman and Smolin offer as a possible solution for the problem
 of time (or, rather, the problem of the apparent absence of time).
 We will begin by arguing that the problem of time, as understood
 by Kauffman and Smolin and perhaps by physicists in general,
 involves a tacit assumption which is both fundamental and
 questionable. We will then show that this assumption is incorrect,
 and that, once it is rejected, a corrected formulation of
 the problem of time will allow for a different solution to
 the problem.
\BP

Kauffman and Smolin, and the other physicists whom they cite,
 appear to understand the problem of the absence of time
 as follows: 
\begin{quote}
If the global time is absent in the sense that
 the constraint equation $Hf = 0$ holds for the total
 wave function $f$ of the universe with $H$ being the
 total Hamiltonian, then we have no time at every
 scale of the universe.
\end{quote}
If the problem is stated in this way, and if we adopt their
definition of local time as the local evolution itself, then the
conclusion means that the local time defined as the evolution
of a local system $L$ cannot exist, and any state $g$ of the local
system $L$ has to satisfy $H_L g = 0$. Here $H_L$ is the local
 Hamiltonian of the local system $L$.

This conclusion --- that the absence of global time implies the
absence of local time --- is the tacit assumption referred to above.
It is not stated explicitly by Kauffman and Smolin. Indeed this
implication does not seem to have been stated explicitly by anybody
in the literature on this problem since it was first raised during
the 1950's. But it is exactly this implication which is responsible
for the feeling that we have a problem of the absence of time at
 {\it every} scale of the universe.
\BP

As a solution of this problem, they appeal to the conjecture that
the total Hilbert space or the total wave function $f$ ``may not be
constructible through any finite procedure" (Abstract of \cite{KS}).
 If this conjecture is true, then they can infer that the
absence of time of the universe cannot be formulated in terms of
any constructible procedure, therefore, the problem of time is ``a
pseudo-problem, because," as they write, ``the argument that time
disappears from the theory depends on constructions that cannot be
realized by any finite beings that live in the universe"
(Abstract of \cite{KS}). Thus ``the whole set up of the problem of time
fails" (p.8, section 3 of \cite{KS}).
\BP

Their argument is thus summarized as follows.
\BP

There appears to be a problem of time because:
\begin{namelist}{88}
\item[1.] If we were to formulate the total wave function for
 the universe, we would find that time was absent at the global
 scale, and 
\item[2.]  The absence of global time implies the absence of
 local time.
\end{namelist}
But:
\begin{namelist}{88}
\item[1a.] The total wave function cannot be constructed by means
 of any finite procedure; hence,
\item[2a.] The absence of global time is never encountered as
 a problem by finite beings, therefore the implication of
 the absence of global time need never be faced by finite beings.
\end{namelist}
In other words, Kauffman and Smolin do not challenge the assumption
that the absence of global time implies the absence of local time.
Their implicit argument is that this assumption is not problematic,
 but just is inconsequential because its condition is never
 instantiated.
\BP

The following argument, however, shows that the assumption that
the absence of global time implies the absence of local time is not
inconsequential, but, on the contrary, has an implication which
seriously undermines the conjecture of Kauffman and Smolin as to the
nature of time.
\BP

Let $B$ and $S$ be two local systems such that $B$
 includes $S$ as a subsystem. The same assumption which
 implies that the absence of global time implies the absence
 of local time would imply in this case that the time $t_S$ of
 the local system $S$ must be equal to the time
 $t_B$ of the bigger system $B$.

If this is the implication of the thought which
 leads to the problem of time, there remains
 a problem in their solution, even if their
 conjecture is true. Consider two disjoint local
 systems $S_1$ and $S_2$, and take a wave function
 $g_j$ of the local system $S_j$  ($j = 1, 2$). Let
 $H_1$ and $H_2$ denote the Hamiltonians associated
 with $S_1$ and $S_2$. Then the time $t_j$ of the system
 $S_j$ is defined by the evolution on the state $g_j$.
 I.e., the local time of $S_j$ is the $t_j$ in the
 exponent of the evolution $\exp[-it_jH_j] g_j$
 (for simplicity, we used this notation to denote
 the path integral which they used in their paper).
 Let $S$ be the union of $S_1$ and $S_2$. Then $S$
 is also a local system, and the tensor product
 $g = g_1 \otimes g_2$ is a wave function of the
 local system $S$. Moreover, the Hamiltonian $H_S$
 of $S$ is given by $H_S = H_1\otimes I + I \otimes H_2$,
 and the time $t_S$ of the system $S$ is given by
 the evolution $\exp[-it_SH_S]g$. According to the
 natural implication mentioned above, we then have
 that $t_j = t_S$ for $j = 1, 2$, because $S$ includes
 both of $S_1$ and $S_2$ as subsystems of $S$.

Therefore, a natural extension of Kauffman and Smolin's argument
yields that all local times of local systems must be identical
 with each other, and there is no local time which is compatible
 with the general theory of relativity.
\BP

Our suggestion is that to remedy this problem it is necessary to
 challenge the assumption that the absence of global time implies
 the absence of local time. As a sufficient basis for that challenge,
 we offer the following observation:

It is a known fact in mathematics, especially known in the area of
mathematical scattering theory, that any local system (consisting of
a finite number of QM particles) can have (internal) motion (i.e.,
can remain an unbound state with respect to the Hamiltonian
associated to the local system), even if the total universe is a
stationary (i.e. bound) state $f$ in the sense that it satisfies the
constraint equation: $Hf = 0$. Namely, this mathematical result means
that the implication mentioned above:
$$
Hf = 0\ {\mbox{implies}}\ H_L g = 0
$$
does not necessarily hold in general.

To explain this result, we take a simple example
 of a local system consisting of 4
 QM particles interacting by electronic Coulomb forces,
 one of which has positive charge, and the other 3
 of which have negative charge. Then it is known
 (H. L. Cycon et al., ``Schr\"odinger Operators,"
 Springer-Verlag, 1987, p. 50) that this system has
 no eigenvalues. Namely, letting $H_3$ denote the
 Hamiltonian of this system, one has that every
 state $g\ne 0$ for this system does not satisfy
 the eigenequation $H_3 g = a g$ for any real number $a$.

However, one can construct a bound system
 by adding one particle with positive charge
 to this system so that one has a system consisting
 of 5 QM particles which has a bound state
 $f\ne 0$ for some eigenvalue $b$. Namely $f$ satisfies
 $H_4 f = b f$ for a real number $b$, where $H_4$
 denotes the Hamiltonian for the extended system
 of 5 particles. Thus we can regard this state $f$
 for this system of 5 particles a mini-universe,
 and the system consisting of 4 particles with
 Hamiltonian $H_3$ introduced at the beginning
 becomes a subsystem of the mini-universe.
 Exactly speaking, the state space of that subsystem
 of the mini-universe $f$ should be the Hilbert space
 $X_3$ which includes all of the state functions
 $f(x,y)$ of the configuration $x$ of the 4 particles,
 with the coordinate $y$ (relative to the center
 of mass of the first 4 particles) of the added
 5th positive charged particle being arbitrary
 but fixed. Thus $X_3$ is a subspace of the Hilbert
 space of all the possible state functions
 for the Hamiltonian $H_3$, and hence,
 by the above-mentioned result of the absence of
 eigenvalues for this Hamiltonian, we have that
 no nonzero state vector in $X_3$ is an eigenstate
 (i.e., bound state) of the Hamiltonian $H_3$.
 Thus the subsystem with Hamiltonian $H_3$ of
 4 particles is not a bound system. I.e.,
 $H_3 g = a g$ does not hold for any real number
 $a$ and any vector $g\ne 0$ in $X_3$.

This example shows that the usual supposition that a subsystem of a
bound system is also a bound system is a mathematically incorrect
statement.
(This argument is a paraphrase of a paragraph of page 8
of \cite{K2}, beginning with ``To state this mathematically, ... .")
\BP

More exactly, we have the following theorem in the context of
 the simplest formulation of quantum mechanics:
\BP

\noindent
{\bf Theorem.}  \ Let $H$ be a $N$-body Hamiltonian with eigenprojection
 $P$ (i.e., the orthogonal projection onto the space of all bound states
 of $H$), with suitable decay assumptions on the pair potentials.
 Let $H$ be decomposed as follows:
$$
H=H_1+I_1+T_1=H_1\otimes I+I_1+I\otimes T_1,
$$
where $H_1$ is a subsystem Hamiltonian with eigenprojection
 $P_1=P_1 \otimes I$ (we use a simplified notation $P_1$ to denote
the extension $P_1 \otimes I$, where $\otimes$ denotes the tensor product
operation and $I$ is an identity operator), $I_1=I_1(x,y)$ is the
 intercluster interaction among the clusters corresponding to the
 decomposition which yields the subsystem Hamiltonian $H_1$, and $T_1$
 is the intercluster free energy.
Then we have
\begin{equation}
(1-P_1)P \ne 0,
\end{equation}
unless the interaction $I_1=I_1(x,y)$ is a constant with respect to $x$
 for any $y$.
\MP

\noindent
{\bf Remark.} In the context of the former part of this paper, this theorem
implies the following: Let $L$ denote a sub local system of an $N$-body
 system with Hamiltonian $H$. Let $H_L$ be the Hamiltonian of that
 local system and let $P_L$ denote the eigenprojection for $H_L$. Then
 the above theorem yields the following:
$$
(1-P_L\otimes I)P \HH_{N} \ne \{ 0 \},
$$
where $\HH_N$ is a Hilbert space of the $N$-body quantum system,
 which could be extended to the Hilbert space of the total universe
in an appropriate sense (see \cite{K1} and \cite{K2}). This relation
 implies that there is some vector $f$ in $\HH_N$ which satisfies
that $Hf=b f$ for some real number $b$ and that $H_L g \ne a g$ for any real
 number $a$, where $g=f(\cdot,y)$ is a state vector of the subsystem
 $L$ with an appropriate choice of the position vector $y$ of
 the subsystem.
\BP

\noindent
{\it Proof} of the theorem. 
Assume that (1) is incorrect. Then we have
$$
(P_1 \otimes I)P=P.
$$
Taking the adjoint operators on the both sides, we also have
$$
P(P_1\otimes I)=P.
$$
Thus $[P_1\otimes I,P] = (P_1\otimes I)P - P(P_1\otimes I) = 0$.
 But in generic this does not hold, because
$$
[H_1,H] = \sum_j^{\mbox{\scriptsize finite sum}} c_j 
\frac{\partial}{\partial x_j}I_1(x,y)\quad (c_j\ \mbox{being
constants})
$$
is not zero unless $I_1(x,y)$ is equal to a constant with respect to $x$.
 Q.E.D.

\BP

Our conclusion is that the absence of global time is compatible with
the existence of local time, and the ``problem of time" as stated by
Kauffman and Smolin and other physicists is not a pseudoproblem, but
an incorrectly formulated problem.
\BP

On the basis of this fact, we can construct the notion of local time
as the evolution associated with each local system, which is proper
 to each local system and is compatible with the general theory of
relativity, without contradicting the nonexistence of global time.
See \cite{K2}, and the references therein.

%\pagebreak

\end{document}